\documentstyle[prl,aps,epsfig]{revtex}

\widetext

\begin{document}

\title{\bf Motional Squashed States}

\author{Stefano Mancini$^{1\,\dag}$,
David Vitali$^{2\,\dag}$, and Paolo Tombesi$^{2\,\dag}$}

\address{
$^{1}$ Dipartimento di Fisica, Universit\`a di Milano,
Via Celoria 16, I-20133 Milano, Italy\\ 
$^{2}$ Dipartimento di Matematica e Fisica, Universit\`a di
Camerino, via Madonna delle Carceri, I-62032 Camerino, Italy \\
$^{\dag}$ Istituto Nazionale per la Fisica della Materia, Italy}

\date{\today}

\maketitle

\begin{abstract}
We show that by using a feedback loop it is possible to
reduce the fluctuations in one quadrature of the vibrational degree
of freedom of a trapped ion below the quantum limit.
The stationary state is not a proper squeezed state, 
but rather a {\it squashed} state,
since the uncertainty in the orthogonal quadrature, which is larger
than the standard quantum limit, 
is unaffected by the feedback action.
\end{abstract}


\section{Introduction}

In recent years there has been an increasing interest on trapping 
phenomena
and related cooling techniques  \cite{pg}.
A number of recent theoretical and experimental papers have investigated
the ability to coherently control or ``engineer" atomic quantum states.
Experiments on trapped ions, where the zero point of motion was closely approached trough laser
cooling \cite{wineprl}, 
already showed the effects of nonclassical motion in the absorption spectrum 
\cite{wineprl,monprl}. 
More recent experiments report the generation of Fock, squeezed and Schr\"odinger cat states 
\cite{winestates,wine}.
These states appear to be of fundamental physical interest and possibly of use for sensitive 
detection of small forces \cite{smallforces}.
Moreover, the possibility to synthetize nonclassical motional states
gave rise to new models
in quantum computation \cite{ciraczoller}.

In this paper we present a
way to reach a stationary nonclassical motional state for a trapped particle \cite{my},
which is able to give a significant uncertainty contraction in one phase-space direction. 
The scheme can be then applied to control the vibrational motion against 
the heating processes 
responsible for decoherence \cite{nist}.
This could be important to obtain high fidelity in quantum logic 
operations \cite{nist}.

The basic idea of the scheme is to realize
an effective and continuous measurement of
a vibrational quadrature for the trapped particle and then apply a feedback
loop able to control, i.e. to reduce, its fluctuations even below the quantum limit. 

The fact that a feedback loop may reduce the fluctuations in one quadrature of an
in-loop field without increasing the fluctuations in the other has been known for a 
long time, and has recently been called ``squashing" \cite{squash} as opposed to ``squeezing"
of a free field, in which the conjugate fluctuations are increased.
In our scheme, the obtained stationary state results as a ``squashed" state;
however, the quadrature measurement
increases the noise in the orthogonal quadrature well above the
standard quantum limit and the squashing comes with 
respect to the state one has in presence of the measurement
process and without the feedback action. In this way the uncertainty principle
is not violated.

The paper is organized as follows. In section II we show
how to realize the indirect continuous measurement of a vibrational quadrature
by coupling the trapped particle with a standing wave. In section III
we shall introduce the feedback loop, in section IV we shall
study the properties of the stationary state in the presence
of feedback and section V is for concluding remarks.

\section{Continuous  monitoring of atomic motion}

We consider a generic particle trapped in an effective harmonic potential.
For simplicity we shall consider the one-dimensional
case, even if the method can be in principle generalized to the
three-dimensional case. This particle can be an ion trapped by a
rf-trap \cite{nist} or a neutral atom in an optical trap \cite{optlat,sara}. 
Our scheme however does not depend on the specific trapping
method employed and therefore we shall always refer from now on
to a generic trapped ``atom''.

The trapped atom of mass $m$, 
oscillating with frequency $\omega_a$ along the
$\hat{x}$ direction and with position
operator $x=x_{0}(a+a^{\dagger})$, $x_{0}=(\hbar/2m\omega_a)^{1/2}$,
is coupled to a standing wave in a cavity with frequency $\omega_{b}$,
wave-vector $k$ along $\hat{x}$ and
annihilation operator $b$. The standing wave is quasi-resonant with the 
transition between two internal atomic levels $|\pm\rangle $
separated by $\hbar\omega_0$.
We consider also an external driving of the standing wave with a 
laser at frequency $\omega_{B}$ and of the atomic center-of-mass motion
with a classical electric field along the $\hat{x}$ direction,
with frequency $\omega_{A}$. The resulting Hamiltonian of the system is
\begin{eqnarray}
&&H=\frac{\hbar \omega_{0}}{2}\sigma_{z} +\hbar\omega_a a^{\dagger}a+
\hbar\omega_{b} b^{\dagger}b +i\hbar \epsilon (\sigma_+ +\sigma_-)
(b-b^{\dagger}) \sin\left(kx+\phi\right) \nonumber \\ 
&&-qEx_0 (a+a^{\dagger})\sin(\omega_A t + \theta)+i\hbar 
\left({\cal B}e^{-i\omega_B t} b^{\dagger} -{\cal B}^{*}
e^{i\omega_B t} b\right) \label{hiniz} \;,
\end{eqnarray}
where $\sigma_z= |+\rangle \langle +|-|-\rangle \langle -|$, 
$\sigma_{\pm}=|\pm \rangle \langle \mp |$, and $\epsilon$
is the coupling constant.

In the interaction representation with respect to
$H_0=\hbar \omega_{B} \left(b^{\dagger} b +\frac{\sigma_{z}}{2}\right)$,
and making the rotating
wave approximation, that is, neglecting terms rapidly
oscillating at the driving laser frequency $\omega_{B}$,
this Hamiltonian becomes
\begin{eqnarray}
&& H=\frac{\hbar \Delta}{2}\sigma_{z} +\hbar\omega_a a^{\dag}a+
\hbar (\omega_b-\omega_B)b^{\dagger}b +i\hbar 
\epsilon (\sigma_+ b - \sigma_-b^{\dagger})
 \sin\left(kx+\phi\right) \nonumber \\
&&  -qEx_0 (a+a^{\dagger})\sin(\omega_A t + \theta)+i\hbar 
\left({\cal B} b^{\dagger} -{\cal B}^{*}
b\right) \;, \label{hrwa}
\end{eqnarray}  
where $\Delta=\omega_0-\omega_B$ is the
atomic detuning. 
This detuning can be set to be much larger than all the other parameters
$\Delta \gg \epsilon$, $\omega_{b}-\omega_{B}$,
and in this case, the excited level can be adiabatically eliminated, 
so to get
the following effective Hamiltonian for the vibrational
motion of the atom and the standing wave mode alone \cite{qo}
\begin{eqnarray}\label{H2}
&& H=\hbar \left(\omega_b-\omega_{B}\right) b^{\dagger}b
+\hbar\omega_a a^{\dagger}a
-\hbar\frac{\epsilon^2}{\Delta} b^{\dag}b\sin^2
\left(kx+\phi\right) \nonumber \\
&&  -qEx_0 (a+a^{\dagger})\sin(\omega_A t + \theta)+i\hbar 
\left({\cal B} b^{\dagger} -{\cal B}^{*}
b\right) \;. \label{hdisp}
\end{eqnarray}

If we set the spatial phase $\phi=0$, and assume the 
Lamb-Dicke regime, one can approximate $\sin^2
\left(kx+\phi\right) \simeq k^2 x^2$ in Eq.~(\ref{hdisp}). Then, in
the interaction representation with respect to $\hbar \omega_A a^{\dagger} a$
and making the rotating wave approximation, i.e., neglecting all the
terms oscillating at $\omega_A$ (which is of the order of 1 Mhz)
or faster because we are
interested in the dynamics at much larger times, we finally get 
\begin{equation}\label{H2b}
H=\hbar \left(\omega_b-\omega_{\cal B}-G/2\right)
b^{\dagger}b
+\hbar(\omega_a-\omega_A) a^{\dagger}a
-\hbar G b^{\dag}b 
a^{\dag} a +i\hbar 
\left({\cal A} a^{\dagger} -{\cal A}^{*}
a\right) +i\hbar 
\left({\cal B} b^{\dagger} -{\cal B}^{*}
b\right) \; ,
\end{equation}
where $G=2(\epsilon k x_0)^2/\Delta$ and $ {\cal A} = -q E e^{-i\theta}/2\hbar$.
This Hamiltonian gives rise
to a crossed Kerr-like effect which could be exploited to
generate nonclassical states analogously to 
the all-optical case proposed in Ref. \cite{kerr}.
A similar approach was used for Schr\"odinger cat 
motional states of atoms in cavity QED \cite{andrew}.
Here, instead, we are looking for stationary
nonclassical states. 
The evolution of the density matrix $D$ of the whole system (vibrational
degree of freedom plus the cavity mode) is determined by the Hamiltonian
(\ref{H2b}) and by the terms describing the photon leakage out of the cavity with decay 
rate $\kappa$ and the coupling of the vibrational motion  
with the thermal environment ${\cal L}_{th}D$, that is
\begin{equation}\label{Dtot}
{\dot D}={\cal L}_{th}D-\frac{i}{\hbar}
\left[H,D\right] 
+\frac{\kappa}{2}\left(2 b D b^{\dag}-b^{\dag} b
D-D b^{\dag} b\right) \;.
\end{equation}
For the determination of the damping term of the vibrational motion
${\cal L}_{th} D$, we note that it occurs at a
frequency $\omega_a $ of the order of MHz 
and that the corresponding damping rate $\gamma$ is 
usually much smaller \cite{wine}. It seems therefore
reasonable to 
use the rotating wave approximation in the interaction
between the atom center-of-mass and its reservoir,
leading us to describe the damping of the vibrational degree of freedom
in terms of the quantum optical master equation (at nonzero
temperature) \cite{qnoise}, 
\begin{equation}
{\cal L}_{th}\rho=\frac{\gamma}{2}(n+1)
\left(2a\rho a^{\dag}-a^{\dag}a\rho-\rho a^{\dag}a\right)
+\frac{\gamma}{2}n
\left(2a^{\dag}\rho a-aa^{\dag}\rho-\rho aa^{\dag}\right)\,,
\label{liu}
\end{equation}
where 
$n=\left[\exp\left(\hbar\nu/k_BT\right)-1\right]^{-1}\,,
$
is the number of thermal phonons ($k_B$ is the Boltzmann 
constant and $T$ the equilibrium
temperature). An analogous treatment is considered in \cite{nist}. 
We have to remark, however,
that the damping and heating mechanisms 
of a trapped atom 
are not yet well understood \cite{nist} and that different kinds of 
ion-reservoir interaction have been proposed \cite{murao}.

The quantum Langevin equations \cite{qnoise} 
corresponding to the master equation
(\ref{Dtot}) reads
\begin{eqnarray}\label{eqsba1}
{\dot b}&=& -i(\omega_b-\omega_{B}-G/2) b+iGa^{\dag} ab
-\frac{\kappa}{2} b+{\cal B}+ \sqrt{\kappa} b_{in}(t)\,,\\
{\dot a}&=& -i(\omega_a-\omega_{A}) a+iGb^{\dag} ba
-\frac{\gamma}{2} a+{\cal A}+\sqrt{\gamma} a_{in}(t)\,,
\label{eqsba2} \end{eqnarray}
where the input quantum noises $b_{in}(t)$ and $a_{in}(t)$ have
zero mean and the following correlation functions
\begin{eqnarray}
&& \langle b_{in}(t)b_{in}(t') \rangle = \langle 
b_{in}^{\dagger}(t)b_{in}(t') \rangle
= 0 \\
&& \langle b_{in}(t)b_{in}^{\dagger}(t') \rangle = \delta(t-t') \\
&& \langle a_{in}(t)a_{in}(t') \rangle =0 \\
&& \langle a_{in}^{\dagger}(t)a_{in}(t') \rangle = 
n\delta(t-t') \\
&& \langle a_{in}(t)a_{in}^{\dagger}(t') \rangle = (n+1)\delta(t-t') \;. 
\end{eqnarray}

When the external driving terms described by ${\cal A}$ and ${\cal B}$
are sufficiently large, the stationary state of the system is quasi-classical,
that is, the standing wave is approximately in a coherent state with a
large amplitude 
$\beta \gg 1$, and the atomic vibrational motion along $\hat{x}$ 
is approximately in a coherent state
with a large amplitude $\alpha \gg 1$. The values of $\alpha$ and 
$\beta$ are given by the solutions
of the coupled nonlinear equations given by the semiclassical version
of the quantum Langevin equations Eqs.~(\ref{eqsba1}) and 
(\ref{eqsba2}):
\begin{eqnarray}
0&=& -i\left(\omega_b-\omega_{B}- G/2-G|\alpha|^2\right)\beta-
\frac{\kappa}{2} \beta+{\cal B}\,,\label{sseqs1}\\
0&=& -i\left(\omega_a-\omega_{A}-G|\beta|^2\right)\alpha-
\frac{\gamma}{2} \alpha+{\cal A}
\,.\label{sseqs2}
\end{eqnarray}
Since $\alpha \simeq 2{\cal A}/\gamma $ and $\beta \simeq 2{\cal 
B}/\kappa $, the semiclassical condition for the steady state is 
satisfied when ${\cal A} \gg \gamma $ and ${\cal B} \gg \kappa$.

The fluctuations around this steady state are instead described by quantum 
mechanics and their dynamics can be obtained by appropriately shifting both
modes, i.e., $b \rightarrow b+ \beta $ and $a \rightarrow a +\alpha $.
In the semiclassical limit $|\alpha |, |\beta | \gg 1$ it is reasonable to 
linearize the equations, and since it is always possible to tune
the two driving frequencies $\omega_A$ and $\omega_B$ so to have zero
detunings, i.e., $|\alpha|^2=(\omega_b-\omega_{B})/G-1/2$,
$|\beta|^2=(\omega_a-\omega_{A})/G$, the linearized quantum Langevin equations
for the quantum fluctuations around the steady state can be written as
\begin{eqnarray}
{\dot b}&=&iG\beta(\alpha^*a+\alpha a^{\dag})-\frac{\kappa}{2} b
+\sqrt{\kappa} b_{in}(t) \,,\label{eqslin1}\\
{\dot a}&=&iG\alpha(\beta^* b+\beta b^{\dag})-\frac{\gamma}{2} a
+\sqrt{\gamma}a_{in}(t) \label{eqslin2}\,.
\end{eqnarray}
The effective linearized Hamiltonian leading to Eqs.(\ref{eqslin1}), (\ref{eqslin2}), 
can be written as
\begin{equation}\label{Heff1}
H=\hbar\chi YX\,,
\end{equation}
where $\chi=-4G|\alpha||\beta |$, 
$Y=(be^{-i\phi_{\beta}}+b^{\dag}e^{i\phi_{\beta}})/2$ 
is the standing wave field quadrature
with phase $\phi_{\beta}$ equal to the phase of the classical amplitude $\beta$,
and $X=(ae^{-i\phi_{\alpha}}+a^{\dag}e^{i\phi_{\alpha}})/2$ is 
the vibrational quadrature
with phase $\phi_{\alpha}$ given by the phase of the classical amplitude $\alpha$.
For the sake of simplicity we shall consider 
$\phi_{\alpha}=\phi_{\beta}=0$, i.e., the atomic position quadrature,
from now on, even if the following considerations can be easily 
extended to the case of generic phases.
Note that in order to remain in the Lamb-Dicke regime 
it is required that $kx_0|\alpha|\ll 1$;
however the linearisation is justified only when $|\alpha|\gg 1$,
and therefore we need $kx_0\ll\frac{1}{|\alpha|}\ll 1$.

Eq.~(\ref{Heff1}) implies that 
an effective continuous, quantum non-demolition (QND)  
measurement of the phonon quadrature $X$ is provided by
the homodyne measurement of the light outgoing from the cavity, which
plays the role of the ``meter''.
In fact, the homodyne photocurrent is \cite{quadra}
\begin{equation}\label{photoc}
I(t)=2\eta\kappa\langle Y_{\varphi}(t)\rangle_c+\sqrt{\eta\kappa} \xi(t)\,,
\end{equation}
where $Y_{\varphi}=(be^{-i\varphi}+b^{\dag}e^{i\varphi})/2$ is
the measured quadrature, the phase
$\varphi$ is related to the local oscillator, and $\eta$ is the 
detection efficiency. The subscript
$c$ in Eq.~(\ref{photoc}) 
denotes the fact that the average is performed on the state 
conditioned on
the results of the previous measurements and
$\xi(t)$ is a Gaussian white noise
\cite{quadra}.
In fact, the continuous monitoring of the field mode
performed through the homodyne measurement,
modifies the time evolution of the whole system. The 
state conditioned on the result of
measurement, described by a stochastic conditioned density matrix 
$D_c$, evolves
according to the following stochastic differential equation 
(considered in the Ito sense)
\begin{eqnarray}\label{Dceq}
{\dot D}_c&=&{\cal L}_{th}D_c-\frac{i}{\hbar}
\left[H,D_c\right] +\frac{\kappa}{2}\left(2 b D_c b^{\dag}-b^{\dag} b
D_c-D_c b^{\dag} b\right)
\nonumber\\
&+&\sqrt{\eta\kappa}\,\xi(t)\left(
e^{-i\varphi} b D_c+e^{i\varphi}D_c b^{\dag}
-2\langle Y_{\varphi}\rangle_cD_c\right)\,.  
\end{eqnarray}
We note that by performing the average over the white noise $\xi(t)$,
one gets the master equation of Eq.~(\ref{Dtot}).

It is now reasonable to assume that the standing wave mode is highly damped,
i.e. $\kappa \gg \chi$ (this does not conflict with the preceding 
assumptions, since the coupling constant 
$\chi = -8 \epsilon^{2}(kx_{0})^{2}|\alpha \beta 
|/\Delta $ is usually smaller than the cavity decay rate).
This means that the radiation field  
will almost always  be 
in its lower state $|0\rangle_b $ (displaced by an amount $\beta$). 
This allows us to adiabatically eliminate the
field and to perform a perturbative calculation in the small 
parameter $\chi /\kappa$, obtaining (see also Ref.~\cite{TV})  
the following expansion for the total conditioned density matrix $D_c$     
\begin{eqnarray}\label{Dofrho}
 D_c&=&\left(\rho_c-\frac{\chi^2}{\kappa ^2}X\rho_c X\right)
\otimes|0\rangle_b{}_b\langle 0|
-i\frac{\chi}{\kappa}\left(
X\rho_c\otimes|1\rangle_b{}_b\langle 0|
-\rho_c X\otimes|0\rangle_b{}_b\langle 1|\right)\nonumber \\
&+& \frac{\chi^2}{\kappa ^2}X\rho_c X \otimes |1\rangle_b{}_b\langle 1|
-  \frac{\chi^2}{\kappa^2 \sqrt{2}}\left(
X^2\rho_c\otimes|2\rangle_b{}_b\langle 0|
+\rho_c X^2\otimes|0\rangle_b{}_b\langle 2|\right) \,,
\end{eqnarray}
where $\rho ={\rm Tr}_{b}\,D$ is the reduced density matrix for 
the vibrational motion. In the adiabatic regime, the internal dynamics
instantaneously follows the vibrational one and therefore
one gets 
information 
on $X$ by observing the quantity $Y_{\varphi}$. 
The relationship between the conditioned mean values follows 
from Eq.~(\ref{Dofrho})
\begin{equation}\label{homoflo}
\langle Y_{\varphi}(t)\rangle_c
=\frac{\chi}{\kappa}\langle X(t)\rangle_c \sin\varphi\,.
\end{equation}

Moreover, if we adopt the perturbative expansion (\ref{Dofrho}) for 
$D_c$ in (\ref{Dceq}) and perform
the trace over the internal mode, we get an equation for the 
reduced density matrix $\rho_c$
conditioned to the result of the measurement of the observable 
$\langle Y_{\varphi}(t)\rangle _c$, and therefore 
$\langle X(t)\rangle _c$
\begin{equation}\label{rhoceq}
{\dot\rho}_c = {\cal L}_{th}\rho_c
-\frac{\chi^2}{2\kappa}
\left[X,\left[X,\rho_c\right]\right]+
\sqrt{\eta\chi^2/\kappa}\,\xi(t)
\left(ie^{i\varphi}\rho_cX-ie^{-i\varphi}X\rho_c
+2\sin\varphi\langle X(t)\rangle_c\,\rho_c\right)\,.
\end{equation}
This equation describes the stochastic evolution of the vibrational state
of the trapped atom conditioned to the result of the continuous
homodyne measurement of the light field. The double commutator
with $X$ is typical of QND
measurements.

\section{The Feedback Loop}

We are now able to use the continuous record of the atom phonon quadrature 
to control its motion through the application of a feedback loop. 
We shall use the continous 
feedback theory proposed by Wiseman and Milburn 
\cite{wisemil}.

One has to take part 
of the stochastic output homodyne 
photocurrent $I(t)$, obtained from the continuous monitoring 
of the meter mode, and 
feed it back to the vibrational dynamics 
(for example as a driving term)
in order to modify the evolution of the mode $a$. 
To be more specific, the presence of feedback modifies the 
evolution of the conditioned state
$\rho_c(t)$. It is reasonable to assume that the feedback 
effect can be described by an additional
term in the master equation, linear in the photocurrent 
$I(t)$, i.e. \cite{wisemil}
\begin{equation}\label{rhofb}
\left[{\dot\rho}_c(t)\right]_{fb}=\frac{I(t-\tau)}{\eta\chi}\,
{\cal K}\rho_c(t)\,,
\end{equation}
where $\tau$ is the time delay in the feedback loop
and ${\cal K}$ 
is a Liouville superoperator describing the way 
in which the feedback signal acts 
on the system of interest.  

The feedback term (\ref{rhofb}) has to be considered in the 
Stratonovich sense, since Eq.
(\ref{rhofb}) is introduced as limit of a real process, then 
it should be transformed in the Ito
sense and added to the evolution equation (\ref{rhoceq}). 
A successive average over the white noise
$\xi(t)$ yields the master equation  for the reduced 
density matrix $\rho={\rm Tr}_{b}D$ 
in the presence of feedback. In the general case of a nonzero 
feedback delay time, one gets a non-Markovian master equation which is 
very difficult to solve \cite{wisemil} (see however Ref.~\cite{delay}).
Most often however, the feedback
delay time is much shorter than the characteristic time of the 
$a$ mode, which in the present case is given by the 
energy relaxation time $\gamma ^{-1}$,
and in this case the dynamics in the presence of feedback can be 
described by a Markovian master
equation \cite{wisemil}, which is given by
\begin{equation}
\dot{\rho }={\cal L}_{th} \rho
 -\frac{\chi^2}{2\kappa}\left[X,\left[X 
,\rho \right]\right]+{\cal K}\left(ie^{i\varphi }\rho X-
ie^{-i\varphi} X\rho\right)
+\frac{{\cal K}^{2}}{2\eta\chi^2/\kappa}\rho.
\label{qndfgen}
\end{equation}
The third term is the feedback term itself and 
the fourth 
term is a diffusion-like term, which is an unavoidable 
consequence of the noise introduced 
by the feedback itself.

Then, 
since the Liouville superoperator ${\cal K}$ can only 
be of Hamiltonian 
form \cite{wisemil}, we choose it as 
${\cal K}\rho =g 
\left[a-a^{\dagger},
\rho \right]/2$ \cite{TV},
which means feeding back the measured homodyne photocurrent to 
the vibrational oscillator with a driving 
term in the Hamiltonian involving the 
quadrature orthogonal to the
measured one; 
$g$ is the feedback gain related to the 
practical way of realizing the loop. 
One could have chosen to feed
the system with a generic phase-dependent quadrature, 
due to the homodyne current, however, it will turn
out that the above choice gives the best and simplest result.
Since the measured quadrature of the vibrational mode is
its position, the feedback will act as a driving for 
the momentum.
Using the above expressions in
Eq.~(\ref{qndfgen}) and rearranging the terms in 
an appropriate 
way, we finally get the following master equation:
\begin{eqnarray}\label{totale}
\dot{\rho }&=& \frac{\Gamma}{2}(N+1)
\left(2a\rho a^{\dagger}-a^{\dagger}a\rho 
-\rho a^{\dagger}a\right)
+\frac{\Gamma}{2}N
\left(2a^{\dagger}\rho a-aa^{\dagger}\rho 
-\rho aa^{\dagger}\right)
\nonumber \\
&-&\frac{\Gamma}{2}M
\left(2a^{\dagger}\rho a^{\dagger}-a^{\dagger 2}\rho 
-\rho a^{\dagger 2}
\right)
-\frac{\Gamma}{2}M^{*}
\left(2a\rho a-a^{2}\rho -\rho a^{2}\right)
\nonumber \\
&-&\frac{g}{4}\sin\varphi
\left[a^{2}-a^{\dag 2},\rho\right]\,,
\end{eqnarray}
where 
\begin{eqnarray}\label{parameters}
\Gamma&=&\gamma-g\sin\varphi\,;\\
N&=&\frac{1}{\Gamma }\left[\gamma n
+\frac{\chi^2 
}{4\kappa}+\frac{g^{2}}{4\eta\chi^2/\kappa}+\frac{g}{2}
\sin\varphi\right]\,;\\
M&=&-\frac{1}{\Gamma }\left[
\frac{\chi^2}{4\kappa}-\frac{g^{2}}{4\eta\chi^2/\kappa}
-i\frac{g}{2}\cos\varphi\right]\,.
\end{eqnarray}
Eq.~(\ref{totale}) is very instructive because 
it clearly shows the 
effects of the feedback loop on the 
vibrational mode $a$. 
The proposed
feedback mechanism, indeed, 
not only introduces a parametric driving term proportional to
$g\sin \varphi$, but it also
simulates the presence of a squeezed bath, 
characterized by an effective damping 
constant $\Gamma $ and by the coefficients
$M$ and $N$, which are given 
in terms of the feedback parameters \cite{TV}. 
An interesting aspect of the effective bath described by 
the first four 
terms in the right hand side of (\ref{totale}) is that 
it is characterized by 
phase-sensitive fluctuations, depending upon the 
experimentally adjustable 
phase $\varphi$.

\section{The Stationary Solution}

Because of its linearity, the solution of  
Eq. (\ref{totale}) can be easily obtained
by using the normally ordered 
characteristic 
function \cite{qnoise} ${\cal C}(\lambda,\lambda^*,t)$.
The partial differential equation corresponding to Eq. 
(\ref{totale}) is
\begin{eqnarray}\label{chareq}
&&\left\{\partial_t+\frac{\Gamma}{2}\lambda\partial_{\lambda}
+\frac{\Gamma}{2}\lambda^*\partial_{\lambda^*}
+\frac{g}{2}\sin\varphi\left(\lambda\partial_{\lambda^*}
+\lambda^*\partial_{\lambda}\right)\right\}{\cal C}(\lambda,
\lambda^*,t)\nonumber\\
&&=\left\{-\Gamma N|\lambda|^2+
\left(\frac{\Gamma}{2}M+\frac{g}{4}\sin\varphi\right)(\lambda^*)^2
+\left(\frac{\Gamma}{2}M^*+\frac{g}{4}\sin\varphi\right)
\lambda^2\right\}
{\cal C}(\lambda,\lambda^*,t)\,,
\end{eqnarray}

The stationary state is reached only if the parameters 
satisfy the stability condition,
i.e. $g\sin\varphi <\gamma $. In this case
the stationary solution has the following form
\begin{equation}\label{charsol}
{\cal C}(\lambda,\lambda^*,\infty)
=\exp\left[-\zeta|\lambda|^2+\frac{1}{2}\mu(\lambda^*)^2
+\frac{1}{2}\mu^*\lambda^2\right]\,,
\end{equation}
where
\begin{eqnarray}
\zeta&=&\frac{N\Gamma^2+g\sin\varphi(\Gamma {\rm Re}\{M\}+2\nu{\rm Im}\{M\})
+g^2\sin^2\varphi/2}
{\Gamma^2-g^2\sin^2\varphi}\,;\label{ze}\\
\mu&=&\Gamma\frac{(N+1/2)g\sin\varphi+\Gamma {\rm Re}\{M\}}
{\Gamma^2-g^2\sin^2\varphi}
+i\frac{(\Gamma^2-g^2\sin^2\varphi){\rm Im}\{M\}} 
{\Gamma^2-g^2\sin^2\varphi}
\label{mu}\,.
\end{eqnarray}

Under the stability conditions and in the long time 
limit $(t\to\infty)$
the variance of the generic quadrature operator
$X_{\theta}=(a e^{i\theta}+a^{\dag}e^{-i\theta})/2$ becomes
\begin{equation}\label{varXth}
\langle X^2_{\theta}\rangle=\frac{1}{2}\left[\frac{1}{2}+\zeta+
{\rm Re}\{\mu e^{2i\theta}\}\right]\,.
\end{equation}

\section{Discussions and Conclusions}

For the $X$ quadrature Eq.(\ref{varXth}) can be simply 
written as
\begin{equation}\label{varXg}
\langle X^2\rangle=\frac{1}{2}\left[\frac{1}{2}+n_{eff}\right]\,,
\end{equation}
with
$
n_{eff}=\zeta+{\rm Re}\{\mu\}$.
In absence of feedback ($g=0$) we have $n_{eff}\equiv n$, 
otherwise
$n_{eff}$ can be smaller than $n$, providing a {\it stochastic
localisation} in the $X$
quadrature. Depending on 
the external parameters,
it can also be negative (but it is always $n_{eff} \ge -1/2$) 
accounting for the possibility of going beyond the standard quantum limit. This is 
a relevant result of the present feedback scheme since it is able to
reduce not only the thermal fluctuations but even the quantum ones.

The potentiality of this feedback mechanism is clearly shown 
in Fig.1, where $n_{eff}$ goes well
below zero for increasing values of $\chi$.
Instead in Fig.2 we have sketched the phase space uncertainty 
contours obtained by cutting the
Wigner function corresponding to Eq.(\ref{charsol}) at 
$1/\sqrt{e}$ times its maximum height. 
We see that the state resulting from the feedback action (solid line) 
has a relevant contraction in the
$X_{\theta=0}$ direction, but the same uncertainty in the 
$X_{\theta=\pi/2}$ direction with respect to the state of the
system undergoing measurement without feedback (dashed line).
We refer to this type of noise reduction produced 
by feedback loop as squashing, whereas squeezing refers 
to conventional quantum noise reduction \cite{squash}.

Summarizing, we have proposed a feedback scheme based 
on an indirect (QND) measurement which is able not only to contain the 
heating of the vibrational motion of a trapped ion, but also to produce nonclassical 
motional states (squashed ones).
Up to this point we have not discussed the specific way
in which a particular feedback
Hamiltonian could be implemented.  
In our case, it is important to be able to realize a term in 
the feedback Hamiltonian
proportional to the quadrature orthogonal to $X$. 
This is not straightforward, but could be realized by using the 
feedback current to vary
an external potential applied to the atom without altering 
the trapping potential \cite{kurt}.
In principle the model could be extended to the three 
dimensional case,
one should only consider three orthogonal standing waves
far from resonant transitions.

In conclusion, although the experimental implementation of the 
presented model may not be easy,  
it is certainly a promising experimental challenge,
stimulated by the possibility of producing 
nonclassical states for trapped atoms
and of controlling their heating to minimize decoherence 
effects, especially in quantum information processing 
\cite{pra}.

\section*{Acknowledgements}

This work has been partially supported by INFM (through the
Advanced Research Project ``CAT"),
by the European Union in the framework of the TMR Network
``Microlasers and Cavity QED" and by MURST under the
``Cofinanziamento".

\bigskip
\bigskip

\bibliographystyle{unsrt}

%\newpage

\begin{figure}[htb]

\centerline{\epsfig{figure=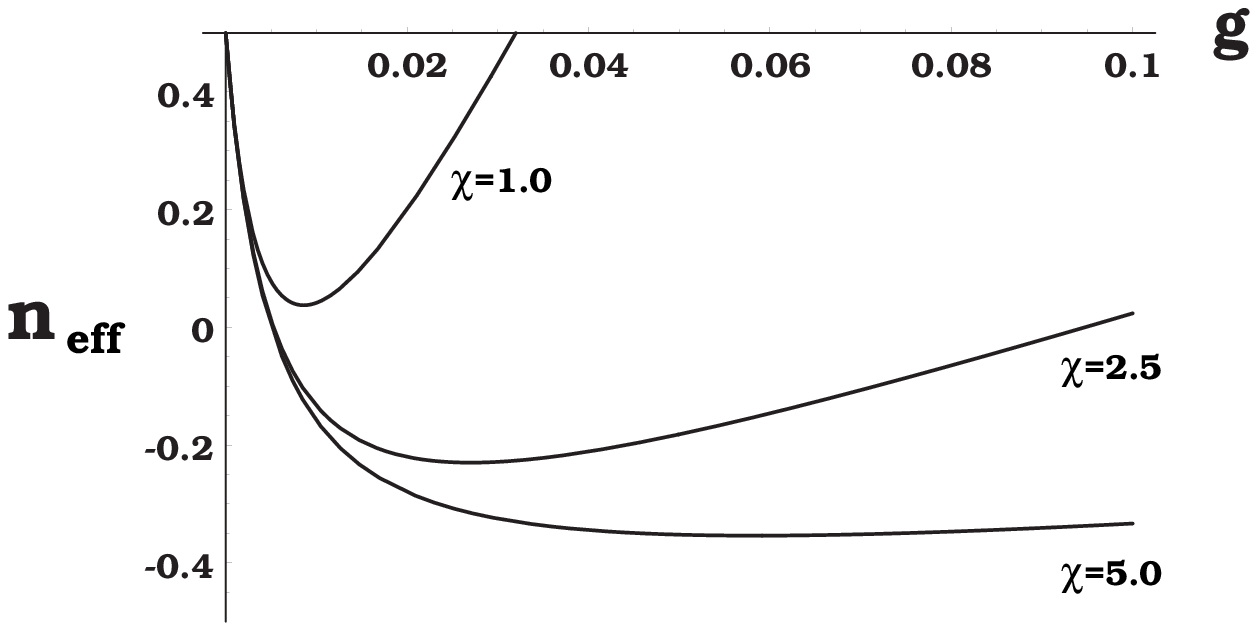,width=8cm}}

\caption{The quantity $n_{eff}$ is 
plotted vs $g$ for different values of $\chi$;
the values of the other parameters are $n=0.5$,
$\gamma=10^{-2}$ ${\rm s}^{-1}$, $\kappa=10^2$ ${\rm s}^{-1}$ and $\eta=0.8$.
The quantities $\chi$ and $g$ are expressed in ${\rm s}^{-1}$.}

\label{fig1}

\end{figure}

\begin{figure}[htb]

\centerline{\epsfig{figure=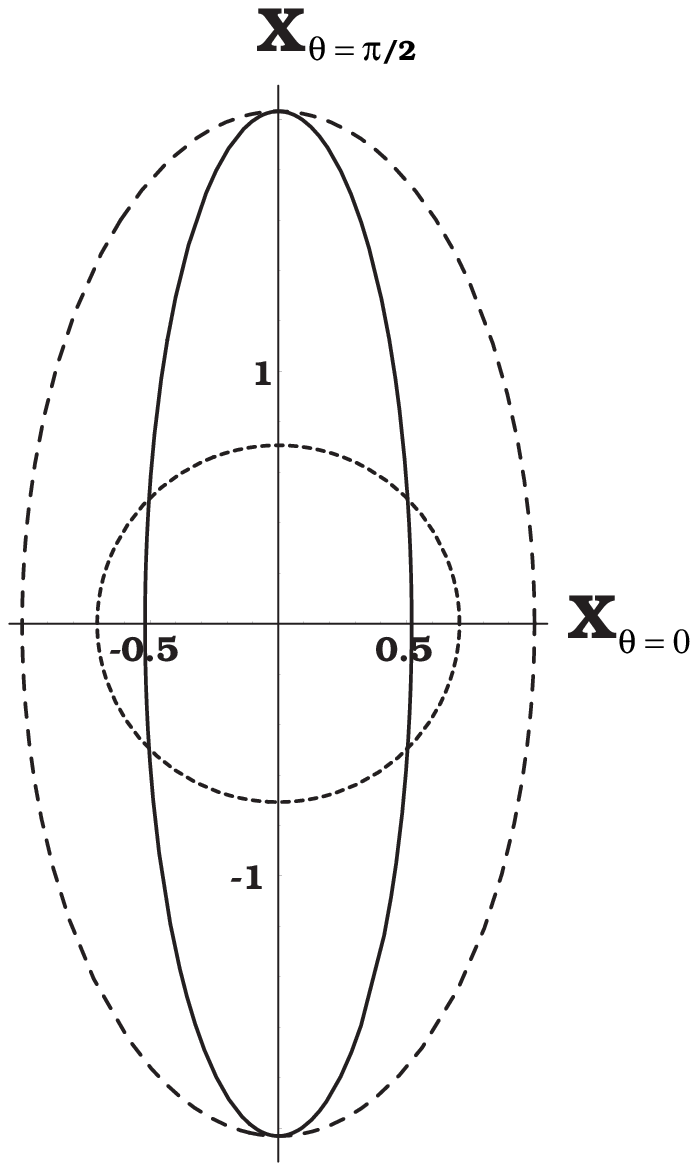,width=8cm}}

\caption{The phase space uncertainty contours are represented for 
$g=0$ (dashed line) and  $g=0.025$ ${\rm s}^{-1}$ (solid line), 
the values of the other parameters are $n=0.5$, $\chi=2.5$ ${\rm s}^{-1}$,
$\gamma=10^{-2}$ ${\rm s}^{-1}$, $\kappa=10^2$ ${\rm s}^{-1}$ and $\eta=0.8$.
The dotted line represents the vacuum noise.}

\label{fig2}

\end{figure}

\end{document}